\documentclass[10pt]{iopart}

\usepackage{amssymb}
\usepackage{graphicx}
\usepackage{hyperref}
\usepackage{bbm}

\newcommand{\eqref}[1]{\eref{#1}}
\newcommand{\eeqref}[1]{equation \eqref{#1}}

\newcommand{\om}{\omega}
\newcommand{\Om}{\Omega}

\newcommand{\stState}{\mathcal{S}}
\newcommand{\pr}[1]{|#1\rangle\langle#1|}
\newcommand{\prt}[2]{|#1\rangle\langle#2|}
\newcommand{\abs}[1]{\left|#1\right|}
\newcommand{\sabs}[1]{|#1|}
\newcommand{\smean}[1]{\langle #1 \rangle}
\newcommand{\ket}[1]{| #1 \rangle}

\newcommand{\mL}{\mathcal{L}}
\newcommand{\mD}{\mathcal{D}}
\newcommand{\mC}{\mathcal{C}}
\newcommand{\mN}{\mathcal{N}}
\newcommand{\mS}{\mathcal{S}}

\newcommand{\id}{\mathbbm{1}}
\newcommand{\ot}{\otimes}

\newcommand{\eff}{{\rm eff}}
\newcommand{\hc}{\textrm{H.c.}}

\begin{document}
\title[Driven-dissipative preparation of entangled states in cascaded quantum-optical \ldots]{Driven-dissipative preparation of entangled states in cascaded quantum-optical networks}
\author{K Stannigel$^{1,2}$, P Rabl$^1$, and P Zoller$^{1,2}$}
\address{$^1$Institute for Quantum Optics and Quantum Information, 
Austrian Academy of Sciences, 6020
Innsbruck, Austria}
\address{$^2$Institute for Theoretical Physics, University of Innsbruck, 6020 Innsbruck, Austria}
\ead{kai.stannigel@uibk.ac.at}
\begin{abstract}
We study the dissipative dynamics and  the formation of entangled states in driven cascaded quantum networks, where multiple systems are coupled to a common unidirectional bath. Specifically, we identify the conditions under which emission and coherent reabsorption of radiation drives the whole network into a pure stationary state with non-trivial quantum correlations between the individual nodes. We illustrate this effect in more detail for the example of cascaded two-level systems, where we present an explicit preparation scheme that allows one to tune the whole network through ``bright" and ``dark" states associated with different multi-partite entanglement patterns. In a complementary setting consisting of cascaded non-linear cavities, we find that two cavity modes can be driven into a non-Gaussian entangled dark state. Potential realizations of such cascaded networks with optical and microwave photons are discussed.
\end{abstract}
\pacs{03.67.Bg,  03.65.Yz, 42.50.Lc}
%\submitto{\NJP}

\maketitle

\section{Introduction}

The coupling of a quantum system to an environment is often associated with decoherence. This is exemplified by the field of quantum computation, where the presence of additional reservoirs degrades the performance of quantum algorithms based on unitary operations executed on large many body systems \cite{Nielsen2000}.
On the other hand, in many quantum optical settings the environment is actually useful for realizing certain applications. Prominent examples are laser cooling or optical pumping, where quantum systems are prepared in highly pure states with the help of a reservoir \cite{Metcalf1999}, or continuous measurement schemes which enable the conditioned preparation of quantum states \cite{Wiseman2010}.
While quantum control schemes employing engineered unitary evolution are by now standard, the last decade has witnessed an increasing interest in alternative methods based on the concept of ``quantum reservoir engineering", i.e., on controlling a quantum system by tailoring its coupling to an environment \cite{Mueller2012}. 
Efforts along these lines have lead, e.g., to a broad range of proposals for the dissipative preparation of entangled few body quantum states \cite{Plenio1999,Parkins2000,Clark2003,Kraus2004,Paternostro2004,Parkins2006,Kastoryano2011}. However, in recent years attention has in particular been devoted to the study of dissipative \emph{many body} systems. 
In this context, it has been realized that quantum reservoir engineering  allows one to dissipatively prepare interesting many body states \cite{Diehl2008,Kraus2008,Verstraete2009,Weimer2010,Cho2011}, perform universal quantum computation \cite{Verstraete2009}, realize a dissipative quantum repeater \cite{Vollbrecht2011}, or a dissipatively protected quantum memory \cite{Pastawski2011}.
Meanwhile, first experiments demonstrating the dissipative preparation of GHZ states  in systems of trapped ions \cite{Barreiro2011} and EPR entangled states of two atomic ensembles \cite{Krauter2011,Muschik2011} have been reported. In these experiments the underlying principle has been to carefully design and implement a many-particle master equation, where a fully dissipative dynamics drives the system into a unique {\em pure} steady state representing the entangled state of interest.

\begin{figure}[t]
\begin{flushright}
\includegraphics[width=0.8\textwidth]{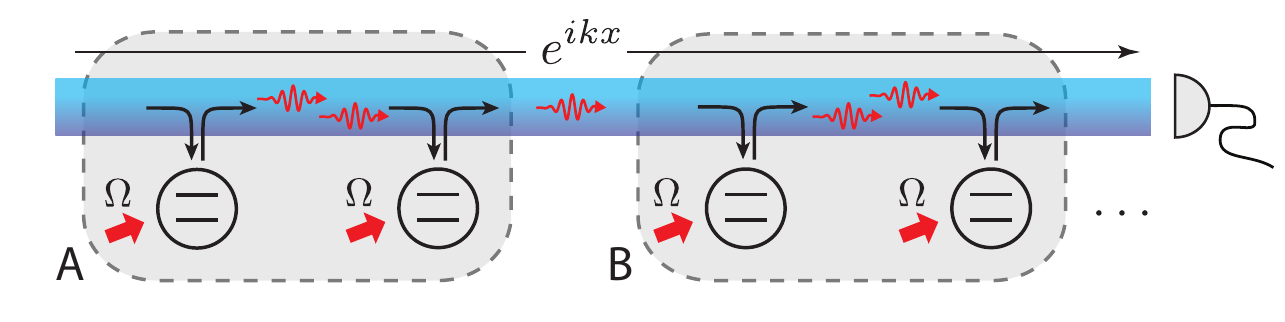}
\end{flushright}
\caption{A driven cascaded quantum network which is realized by a set of two-level systems coupled to a unidirectional bath.
The two-level systems are driven by classical fields $\sim \Omega$ and the continuously emitted radiation propagates along the waveguide and excites successive nodes.  Under specific conditions, all photons emitted in a subsystem A are coherently reabsorbed in subsystem B.  In this case the system relaxes into a \emph{dark state} where no radiation escapes from the network, but a constant stream of photons running from  A to B establishes entanglement between the two subsystems.}
\label{fig:figure1}
\end{figure}

In this work, we study entanglement formation in driven few- and many-particle {\em cascaded quantum networks} as introduced by Gardiner and Carmichael \cite{Gardiner1993,Carmichael1993}, where the unconventional coupling of multiple systems to a common unidirectional bath offers remarkable new opportunities for  dissipative preparation of highly correlated states.
As illustrated in \fref{fig:figure1}, a cascaded quantum network consists of $N$ systems coupled to a 1D reservoir that has the unique feature that excitations can only propagate along a \emph{single} direction, thereby driving successive systems in the network in a unidirectional way. 
While such a scenario is reminiscent of edge modes in quantum Hall systems \cite{Kane1997,Haldane2008,Wang2008,Wang2009,Hafezi2011}, various artificial and more controlled realizations based on 
(integrated) non-reciprocal devices for optical \cite{Wang2005,Feng2011} and microwave \cite{Koch2010,Kamal2011} photons are currently developed.
Our goal below is to identify situations where cascaded quantum networks are driven by classical fields in such a way that they exhibit pure and entangled steady states. As shown in \fref{fig:figure1}, such \emph{dark states} emerge if  the continuous stream of photons emitted by the first part (``A") of the network is coherently reabsorbed by the second part (``B"), such that no photons escape from the system and the output remains dark. The system hence acts as its own {\em coherent quantum absorber} while the constant stream of photons maintains entanglement between its two parts.

%%%%%%%%%%%%%%%%%%%%%%%%%%%%%%%%%%%
%%%           FIGURE            %%%
%%%%%%%%%%%%%%%%%%%%%%%%%%%%%%%%%%%

We will first discuss the construction of coherent quantum absorbers on general grounds. Given the first part A of the network, we show how to choose the second part B such that the whole system evolves into a dark state. In doing so, we make use of the unidirectionality of the reservoir, which allows us to solve the first part independently of the second. These formal developments are then illustrated by two settings in which the coherent quantum absorber scenario can be realized. The first is a many body cascaded network, where each of the $N$ nodes consists of a driven two-level system (TLS, ``spin"). We show that this system exhibits a whole class of multi-partite entangled dark states, whose entanglement structure can be adjusted by tuning local parameters. As a second, complementary system we consider a network consisting of two non-linear cavities described by bosonic mode operators. By choosing appropriate laser drives  and Kerr-type non-linearities one can ensure that this system evolves into a non-Gaussian dark state in which the two modes are entangled. In both examples, the coherent quantum absorber scenario thus leads to a dissipative state preparation scheme for non-trivial entangled states. In a more general context, these networks realize a novel type of non-equilibrium (many body) quantum system, which by changing the system parameters can be tuned between ``dark" or ``passive" phases (with no scattered photons emerging) and ``bright" or ``active" phases (with light scattered), while the nodes are driven into pure entangled or mixed states, respectively.

The remainder of this paper is structured as follows. In section \ref{sec:absorbers} we present the general model for an $N$-node cascaded network and show how to construct the coherent quantum absorber subsystem B for some given subsystem A (cf. \fref{fig:figure1}). Complementing these rather general developments, we discuss the two mentioned realizations of the coherent quantum absorber scenario in subsequent sections.  Section \ref{sec:spin} presents the driven cascaded spin-system for which we derive and discuss a multi-partite entangled class of dark states and also comment on the influence of various imperfections. Subsequently, section \ref{sec:kerr}  discusses the setup based on cascaded  cavities with Kerr-type non-linearity. In the latter case, we also obtain a purification of the steady state density matrix of the well-known dispersive optical bi-stability problem \cite{Drummond1980} as an interesting by-product. Implementations of the proposed cascaded networks are discussed in \sref{sec:implementations} and concluding remarks can be found in \sref{sec:conclusion}.

\section{Coherent quantum absorbers}
\label{sec:absorbers}

\subsection{Cascaded quantum networks} 
\label{sec:cascadedSystems}

We consider the general setting of a cascaded quantum network as shown in \fref{fig:figure1}.
Here, $N\geq 2$ subsystems located at positions $x_i$ are coupled to a 1D continuum of right-propagating bosonic modes $b_\om$, which represent, for  example, photons in an optical or microwave waveguide. 
The whole network can be modeled by a Hamiltonian  ($\hbar=1$)
\begin{eqnarray} \label{eq:Htot}
H=
\sum_i H_i +  H_{\rm bath} 
+ \sum_i \int \rmd\om\,  g_\omega \left( c_i^\dag  b_{\om} \rme^{\rmi \om x_i/v}  + \hc \right),
\end{eqnarray}
where $H_{\rm bath}=\int\rmd\om\,\om\,  b_\om^\dag b_\om$ is the free Hamiltonian of the bath modes and the integrals run over a broad bandwidth $\Delta \omega$ around the characteristic system frequency $\om_0$. In this frequency range the bath modes are assumed to exhibit a linear dispersion relation with speed of light $v$. In \eeqref{eq:Htot} the $H_i$ describe the dynamics of the individual systems and include classical driving fields (e.g. lasers or microwave fields). The system-bath coupling is determined by the ``jump operators" $c_i$ and coupling constants $g_{\om}$, which we assume to be approximately constant over the frequency range $\Delta \omega$. 
Implementations of an effective model of the form given in \eeqref{eq:Htot} can be achieved with atoms or solid state TLSs coupled to 1D optical or microwave waveguides and will be discussed in more detail  in \sref{sec:implementations} below.

The system-bath interaction in \eeqref{eq:Htot} breaks time reversal symmetry and while photons can be emitted to the right, drive successive subsystems and eventually leave the network, the reverse processes cannot occur.  To study the effects of this unconventional coupling, we assume that $\om_0$ as well as the bandwidth $\Delta \omega$ are large compared to the other relevant frequency scales and eliminate the bath modes in a Born-Markov approximation.
This yields a generalized cascaded master equation (ME) for the 
reduced system density operator $\rho$~\cite{Gardiner1993,Carmichael1993,Stannigel2011},
\begin{equation}\label{eq:ME1}
%\begin{split}
\dot \rho= \sum_i \mathcal{L}_i \rho
 -\gamma \sum_{j>i}\left([c_j^\dag,c_i \rho] +  [\rho c_i^\dag,c_j]  \right).
%\end{split}
\end{equation}
Here the first part describes the uncoupled evolution of each subsystem $\mathcal{L}_i \rho= -i[H_i,\rho] +\gamma \mathcal{D}[c_i]\rho$, where the Lindblad terms 
$\mD[x]\rho=x \rho x^\dag-\{x^\dag x,\rho\}/2$ model dissipation due to emission of photons into the waveguide with a rate $\gamma=2\pi g_{\om_0}^2$. The unidirectionality of the bath is reflected by the last term in \eeqref{eq:ME1}, which accounts for the possibility to reabsorb photons emitted at system $i$ by all successive nodes located at $x_j>x_i$.  
The explicit Lindblad form of the ME\,\eqref{eq:ME1} reads
\begin{equation}\label{eq:ME2}
\dot \rho = -\rmi [H_{\rm casc}, \rho] +\gamma \mathcal{D}[c]\rho\,,
\end{equation} 
where $H_{\rm casc}= \sum_i H_i - \rmi\frac{\gamma}{2}\sum_{j>i} (c_j^\dag c_i  -  c_i^\dag c_j)$ now includes the non-local coherent part of the environment-mediated coupling, while the only  decay channel with collective jump operator $c=\sum_ic_i$ is associated with a photon leaving the system to the right.\footnote{For $N=2$ this ME stands in contrast to Ref.~\cite{Krauter2011,Muschik2011}, where the dynamics is purely dissipative with no coherent evolution.} Note that equations \eqref{eq:ME1} and \eqref{eq:ME2} are understood in a rotating frame in order to account for the explicit time-dependence of the driving fields.

\subsection{Coherent quantum absorbers} 
\label{sec:constructionAbsorbers}

In the following we are interested in steady state situations where every photon emitted within the system is perfectly reabsorbed by successive nodes in the network, 
such that there is no spontaneous emission via the waveguide output and the system relaxes to a pure steady state $\rho_0=|\psi_0\rangle\langle \psi_0|$.
To identify the general conditions for the existence of such states, we partition the network into two subsystems A and B as indicated in \fref{fig:figure1},
with local Hamiltonians $H_A$ and $H_B$, and jump operators $c_A$ and $c_B$, respectively. Specifically,  $H_A= \sum^\prime_i H_i - \rmi\frac{\gamma}{2}\sum^\prime_{j>i} (c_j^\dag c_i  -  c_i^\dag c_j)$ and $c_A= \sum^\prime_i c_i$, where the primed sums run over the nodes of part A, and corresponding expressions hold for B. 
Then, in \eeqref{eq:ME2}, 
$H_{\rm casc}= H_A+H_B- \rmi \frac{\gamma}{2}(c_A c_B^\dag  - c^\dag_A c_B )$ and $c=c_A+c_B$,  and the conditions for the existence of a pure stationary state are (see Ref.\,\cite{Kraus2008} and further comments in \ref{sec:app:absorber}):
\begin{equation} 
\label{eq:conditions}
{\bf (I)}\,\, \, (c_A+c_B)|\psi_0\rangle =0\,, \qquad  
{\bf (II)}\,\,\,  [H_{\rm casc},\rho_0]=0\,.
\end{equation}
The first condition implies that the waveguide output is dark, i.e., $\smean{c^\dag c}=0$, and the second one ensures stationarity. Within a quantum trajectory picture, condition ${\bf (I)}$ means that there are no stochastic quantum jumps, which would lead to a mixed state.
In situations where $[H_{A,B},\rho_0]=0$ 
and $c_{A,B}|\psi_0\rangle=0$ for each subsystem separately, the network can simply be devided into two smaller parts which are then treated independently.
In the following, we thus focus on situations where such a division is not possible and where the steady state possesses non-trivial correlations between A and B, as characterized for example by a non-vanishing $\mathcal{C}\equiv\langle c_A^\dag c_B+c_A c_B^\dag \rangle - 2 {\rm Re} \{ \langle c_A^\dag \rangle\langle c_B\rangle\}$.
In view of ${\bf (I)}$ this third requirement can be expressed as
\begin{equation}
{\bf (III)}\,\, \,  \mathcal{C}= - 2 (\langle c_A^\dag c_A\rangle - |\langle c_A \rangle|^2)\neq  0\,,
\end{equation}
and directly connects the correlations between A and B with the amount of radiation emitted from the first subsystem.  Note that for a pure steady state, $\mC\neq 0$ also implies that A and B are entangled. Finally, we remark that under stationary conditions, a non-vanishing $\mathcal{C}$ implies a constant flow of energy from A to B, while the total scattered light vanishes. In the examples discussed below this `coherent' absorption of energy in  B can be understood as a destructive interference of the signals scattered by the two subsystems.

The conditions ${\bf (I)}-{\bf (III)}$ will not be satisfied in general. However, given a system A described by a Hamiltonian $H_A$ and jump operator $c_A$ we can construct a perfect coherent absorber system B as follows. 
First, we point out that due to the unidirectional coupling
the dynamics of A is unaffected by B, which can also be shown explicitly by tracing the ME \eref{eq:ME2} over system B. In particular, the steady state $\rho_A^0$ of A is obtained by solving $\mathcal{L}_A\rho_A^0=0$, where 
$\mL_A\rho_A \equiv -\rmi[H_A,\rho_A] + \gamma\mD[c_A]\rho_A$, and  assuming a unique solution we write its spectral decomposition as $\rho_A^0= \sum_k p_k |k\rangle\langle k|$.
A pure state of the whole system is then given by $|\psi_0\rangle=\sum_k \sqrt{p_k} |k\rangle_A\otimes |\tilde k\rangle_B$, where we assumed A and B to have the same Hilbert space dimension and defined $|\tilde k\rangle= V|k\rangle$ in terms of an arbitrary unitary $V$ acting on $\ket{k}$.
 Now, we demonstrate in \ref{sec:app:absorber} that conditions ${\bf (I)}$ and ${\bf (II)}$ can be satisfied by the choice
\begin{eqnarray}
\label{eq:cB}
c_B&= - \sum_{n,m}  \sqrt{\frac{p_n}{p_m}}  \langle m| c_A| n\rangle  |\tilde n\rangle\langle \tilde m|_B\,,\\
\label{eq:HB}
H_B&= - \frac{1}{2} \sum_{n,m}  \left( \sqrt{\frac{p_n}{p_m}} \,A_{mn}+\sqrt{\frac{p_m}{p_n}} \,A_{nm}^* \right)   |\tilde n\rangle\langle \tilde m|_B\,,
\end{eqnarray}
where $A_{mn}= \langle m| H_{A,{\rm eff}}| n\rangle $
and  $H_{A,{\rm eff}}=H_{A}-\rmi\frac{\gamma}{2}c_A^\dag c_A$ is the effective non-hermitian Hamiltonian associated with $\mL_A$, and we assumed a positive and non-degenerate spectrum $\{p_k\}$.
While equations \eqref{eq:cB} and \eqref{eq:HB} define a general absorber system B, we find that for many systems of interest the stationary state $\rho_A^0$ satisfies $\sqrt{p_n} \langle k | c_A|n\rangle =\sqrt{p_k} \langle n | c_A|k\rangle $ and $\sqrt{p_n} \langle k | H_{A,{\rm eff}}|n\rangle =\sqrt{p_k} \langle n | H_{A,{\rm eff}}|k\rangle $.  
In this case, the above relations simplify to 
$c_B=-Vc_AV^\dag $ and $H_B=-VH_AV^\dag$ \footnote{We use this as a short-hand notation for an equivalent matrix representation of the two operators,  $\langle n| c_B|m\rangle =-  \langle n| Vc_AV^\dag |m\rangle$, etc. Here, the same set of states is used on both systems, which is permissible due to their equal Hilbert-space dimension.}
such that  up to a unitary basis transformation the absorber system is just the negative counterpart of A. In particular, this situation applies to the two examples of cascaded spin systems and cascaded non-linear cavities, which we describe in more detail in the following sections.

\section{Cascaded spin networks} 
\label{sec:spin}

Let us now be more specific  and consider a set of $N$ driven spins coupled to a unidirectional bosonic bath as shown in \fref{fig:figure1}. 
In \eeqref{eq:ME2}, the collective jump operator is now $c=\sum_i\sigma_-^i$ and the cascaded Hamiltonian in the frame rotating at the frequency $\om_d$ of the external driving field reads
\begin{equation}
\label{eq:Hcasc}
H_{\rm casc}= \sum_{i}\left(\frac{\delta_i}{2} \sigma_z^i+ \Omega_i \sigma_x^i\right) 
- \rmi\frac{\gamma}{2}\sum_{j>i}\left(\sigma_+^j\sigma_-^i - \sigma_-^j\sigma_+^i\right).
\end{equation}
Here the $\sigma_\mu^i$ are the usual Pauli operators on site $i$, the $\Om_i$ are local Rabi frequencies, and the $\delta_i=\om_i-\om_d$ are the detunings of the individual spin transition frequencies $\om_i$ from the common classical driving frequency $\omega_d$. In the following, the basis states of the spins are denoted by $\ket{e}$, $\ket{g}$, such that $\sigma_-=\prt{g}{e}$. The classical fields which are used to drive the spins can either be applied via additional local channels (assuming that the associated decay rate is much smaller than $\gamma$) or via a coherent field which is sent through the common waveguide. Note that by omitting the cascaded interaction in equation \eref{eq:Hcasc} we recover the familiar Dicke model for multiple two level atoms decaying via a common field mode, where even for $\Omega=0$ a series of  dark states can be identified from $c|\psi_0\rangle=0$. In contrast, in our cascaded setting these states are not stationary and non-trivial dark states can only emerge from an interplay between driving and cascaded coupling terms.

\subsection{Construction of dark states}

For $N=2$ the dark state condition ({\bf I}) restricts $|\psi_0\rangle$  to the subspace spanned by $|gg\rangle$ and the singlet $\ket{S}=(\ket{eg}-\ket{ge})/\sqrt{2}$. Condition ({\bf II}) can then be satisfied for $\Omega_1=\Omega_2\equiv\Omega$ and any $\delta_1=-\delta_2\equiv\delta$, for which we obtain the unique and pure steady state $|\psi_0\rangle=|\stState_\delta\rangle$, where 
\begin{eqnarray}\label{eq:S2}
\ket{ \stState_\delta}=\frac{1}{\sqrt{1+\sabs{\alpha}^2}}\left(\ket{gg}+\alpha \ket{S}\right),
\qquad
\alpha=\frac{2\sqrt{2}\Om}{\rmi\gamma-2\delta}\,.
\end{eqnarray}
The two spins thus realize a source and a matched absorber in the sense introduced in the previous section, and for the matrix representations of the various operators we can identify 
$c_A=\sigma_-$, $c_B=-V c_A V^\dag=\sigma_-$ and $H_B=-VH_AV^\dag$, using $V=\sigma_z$.  
For strong driving, $\abs{\alpha}\gg1$, the state $\ket{\stState_\delta}$ approaches the singlet $\ket{S}$, where the mutual correlation $\sabs{\mC}\rightarrow 1$ is maximized.

While for larger $N$ a direct search for possible dark states is hindered by the exponential growth of the subspace defined by $c|\psi_0\rangle=0$,
we can use the state \eqref{eq:S2} as a starting point and solve the cascaded system iteratively ``from left to right": 
Suppose that for $\Omega_i=\Omega$ and $\delta_1=-\delta_2$ the first two spins have evolved into the dark state $\ket{\stState_{\delta_1}}$
such that no more photons are emitted into the waveguide. Then, the following two spins effectively see an empty waveguide and evolve into the dark state $\ket{\stState_{\delta_3}}$, provided that $\delta_3=-\delta_4$. By iterating this argument we see that for any detuning profile with $\delta_{2i-1}=-\delta_{2i}$ ($i=1,2,\ldots$) the steady state of the ME \eqref{eq:ME2} is given by
\begin{eqnarray}
\label{eq:psi0}
\ket{\stState^0}
=\ket{\stState_{\delta_1}}_{12} \otimes \ket{\stState_{\delta_3}}_{34}\otimes \ldots \,,
\end{eqnarray}
which is shown explicitly in \ref{sec:app:spin}.
In particular, for the homogenous case $\delta_i\simeq 0$, a strongly driven cascaded spin system relaxes into a chain of pairwise singlets.

%%%%%%%%%%%%%%%%%%%%%%%%%%%%%%%%%%%
%%%           FIGURE            %%%
%%%%%%%%%%%%%%%%%%%%%%%%%%%%%%%%%%%
\begin{figure}
\begin{flushright}
\includegraphics[width=0.8\textwidth]{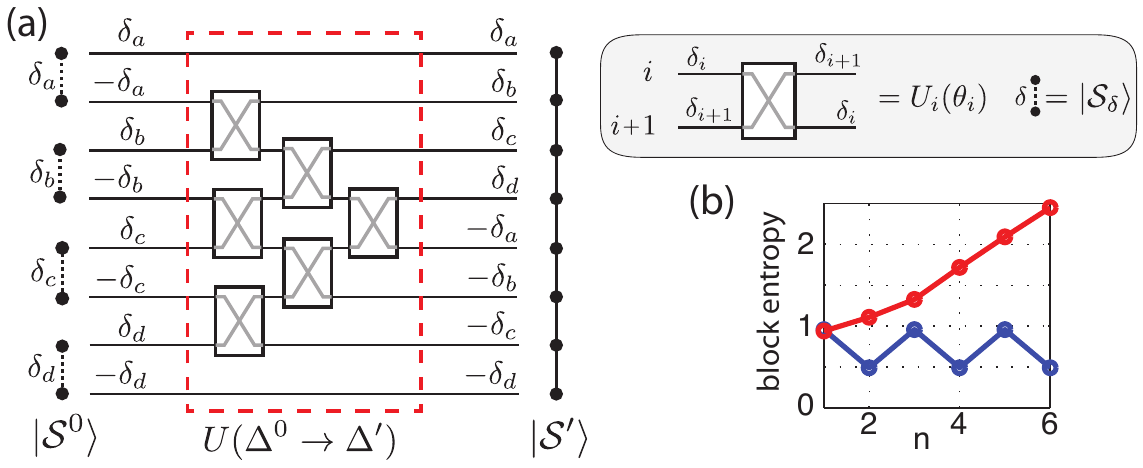}
\end{flushright}
\caption{(a) Circuit model for constructing the state $|\stState^\prime\rangle$ in \eeqref{eq:classOfDarkStates} for the example $\Delta^\prime=(\delta_a,\delta_b,\ldots,-\delta_a,-\delta_b,\ldots)$.
Each line represents a spin and is labeled by its detuning. 
A box connecting two lines denotes a unitary operation $U_i(\theta_i)$ and the corresponding exchange of detunings $\delta_i$ and $\delta_{i+1}$ (see text and inset).
(b) Scaling of the von-Neumann entropy of the first $n$ spins in a network of $N=12$ nodes for $\Omega=2\gamma$. Upper curve: detuning profile as in (a) with $\sabs{\delta_i}=\gamma/3$. Lower curve: state corresponding to the detuning profile $\Delta^\prime=(0,\gamma,-\gamma,\gamma,\ldots,-\gamma,0)/3$.}
\label{fig:MultiQubitStates}
\end{figure}

The dimer structure of the state $\ket{\stState^0}$ reflects the fact that radiation emitted from one node is immediately reabsorbed by the following one.
We now consider situations where this reabsorption occurs by several of the following spins,  leading to multi-partite entangled states.
To this end, note that by starting from the state in \eeqref{eq:psi0} we can construct another dark state $|\stState^\prime\rangle=U\ket{\stState^0}$ by any global unitary operation $U$ with $[U,c]=0$, while implementing the Hamiltonian $H'=U H_{\rm casc} U^\dag$ would ensure stationarity.  
However, for arbitrary unitary transformations $H^\prime$ will in general contain additional non-local terms, and we must hence restrict ourselves to unitaries $U$ under which $H_{\rm casc}$ is \emph{form invariant}. 
As an example, we  write $H_{\rm casc}\equiv H_{\rm casc}(\Delta)$, where $\Delta=(\delta_1,\delta_2,...)$ is the detuning profile,  and introduce the nearest-neighbor operations
\begin{equation}\label{eq:U}
U_{i}(\theta_i)=\exp\left[\rmi\frac{\theta_i}{4}\left(\vec{\sigma}_i+\vec{\sigma}_{i+1}\right)^2\right]
\propto \exp\left[\rmi\frac{\theta_i}{2}\,\vec{\sigma}_i\cdot\vec{\sigma}_{i+1}\right]\,,
\end{equation}
where $\vec{\sigma}_i=(\sigma_x^i,\sigma_y^i,\sigma_z^i)$.
Then, by choosing 
$\tan(\theta_i)=(\delta_{i+1}-\delta_i)/\gamma$ we obtain 
\begin{equation}\label{eq:formInvariance}
H'=U_i(\theta_i)\, H_{\rm casc} (\Delta) \, U_i^\dag(\theta_i) = H_{\rm casc} (\Delta^\prime)\,,
\end{equation}
with a new detuning profile $\Delta^\prime=P_{i,i+1}\Delta$, where $P_{i,i+1}$ denotes the permutation of $\delta_i$ and $\delta_{i+1}$ (this is demonstrated in \ref{sec:app:spin}). Thus, by starting from a set $\Delta^0$ of alternating detunings as defined before \eeqref{eq:psi0}, we can simply swap the detunings of nodes $i$ and $i+1$ to implement a new cascaded spin network with a unique stationary state $|\stState^\prime\rangle=U_i(\theta_i)\ket{\stState^0}$. By repeating this argument, we obtain a different pure steady state for each permutation $\Delta^\prime$ of  $\Delta^0$. This class of states is given by 
\begin{eqnarray}
\label{eq:classOfDarkStates}
\ket{\stState ^\prime}=U(\Delta^0\rightarrow \Delta^\prime)\ket{\stState^0}\,,
\end{eqnarray}
where $U(\Delta^0\rightarrow \Delta^\prime)$ is a product of nearest-neighbor operations $U_i(\theta_i)$, specified by the sequence of nearest-neighbor transpositions required for transforming $\Delta^0$ into $\Delta^\prime$. A graphical representation of $U(\Delta^0\rightarrow \Delta^\prime)$ in terms of a circuit model is shown in \fref{fig:MultiQubitStates}(a) for a specific example.

%%%%%%%%%%%%%%%%%%%%%%%%%%%%%%%%%%%
%%%           FIGURE            %%%
%%%%%%%%%%%%%%%%%%%%%%%%%%%%%%%%%%%
\begin{figure}
\begin{flushright}
\includegraphics[width=0.8\textwidth]{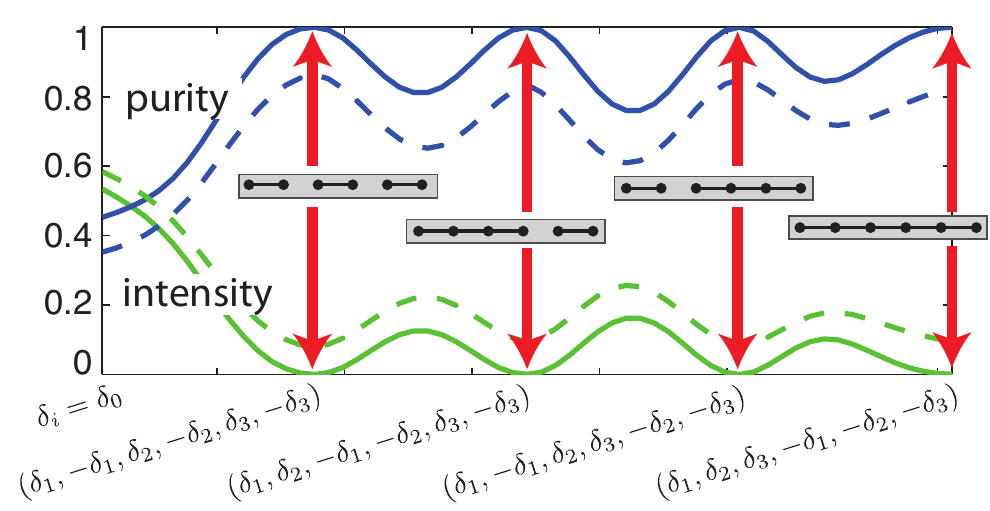}
\end{flushright}
\caption{Tuning a six-spin network through bright and dark states by adjusting local detunings. 
The curves show purity and output intensity $\smean{c^\dag c}$, where the  detunings $\delta_i$ are interpolated linearly between the profiles given on the horizontal axis. Red arrows indicate the entanglement structure of the dark states. The solid lines show the ideal case, while for the dashed lines we included a finite on-site decay $\kappa_0=0.0025\gamma$ (see \sref{sec:imperfections}). Parameters are  $\Om=\gamma$, $\delta_0=\gamma/5$, $\delta_1=\gamma$, $\delta_2=\gamma/2$, $\delta_3=0$.
}
\label{fig:brightAndDarkStates}
\end{figure}

\subsection{Discussion} 

For large detuning differences $|\delta_i-\delta_{i+1}|\gg\gamma$ the unitary transformations given in \eeqref{eq:U} are SWAP operations \cite{Nielsen2000} between neighboring sites. In this limit  the states $\ket{\stState^\prime}$ remain approximately two-partite entangled,
but with singlets shared between arbitrary nodes in the network. 
In contrast, for $|\delta_i-\delta_{i+1}| \approx \gamma$ the $U_i$ correspond to highly entangling $\sqrt{{\rm SWAP}}$ operations. Then, the entanglement structure can be much richer and in general the states $\ket{\stState^\prime}$ contain multi-partite entanglement between several or even all nodes.  While in this case a full characterization is difficult, we point out that the $U_i$ conserve total angular momentum such that the $\ket{\stState^\prime}$ approach multi-spin singlets in the strong driving limit.
The amount of entanglement between subsystems now depends very much on the choice of the detuning profile, as can be seen from the two examples displayed in \fref{fig:MultiQubitStates}(b), showing oscillating and linearly growing block entropy, respectively. More generally, the cascaded network can be driven into different types of states by simply adjusting local detunings.
This is illustrated in \fref{fig:brightAndDarkStates}, where an adiabatic variation of the detunings in a six-node network is used to prepare pure steady states with 2-, 4- and 6-partite entanglement, separated by ``bright" (mixed state) phases where the conditions in \eeqref{eq:conditions} are violated.

For assessing the relaxation time that characterizes the approach of the system towards the presented steady states it is sufficient to examine the simple detuning profile leading to $\ket{\mS^0}$, since the spectral properties of the Liouvillian are invariant under the transformations \eref{eq:U}. Numerical calculations for small systems suggest that the preparation time for these states scales efficiently with the number of nodes $N$. In particular, the uniqueness of $\ket{\mS^0}$ implies that effects related to non-unique steady states found in related systems \cite{Gu2006} are absent in our case.

%%%%%%%%%%%%%%%%%%%%%%%%%%%%%%%%%%%
%%%           FIGURE            %%%
%%%%%%%%%%%%%%%%%%%%%%%%%%%%%%%%%%%
\begin{figure}
\begin{flushright}
\includegraphics[width=0.8\textwidth]{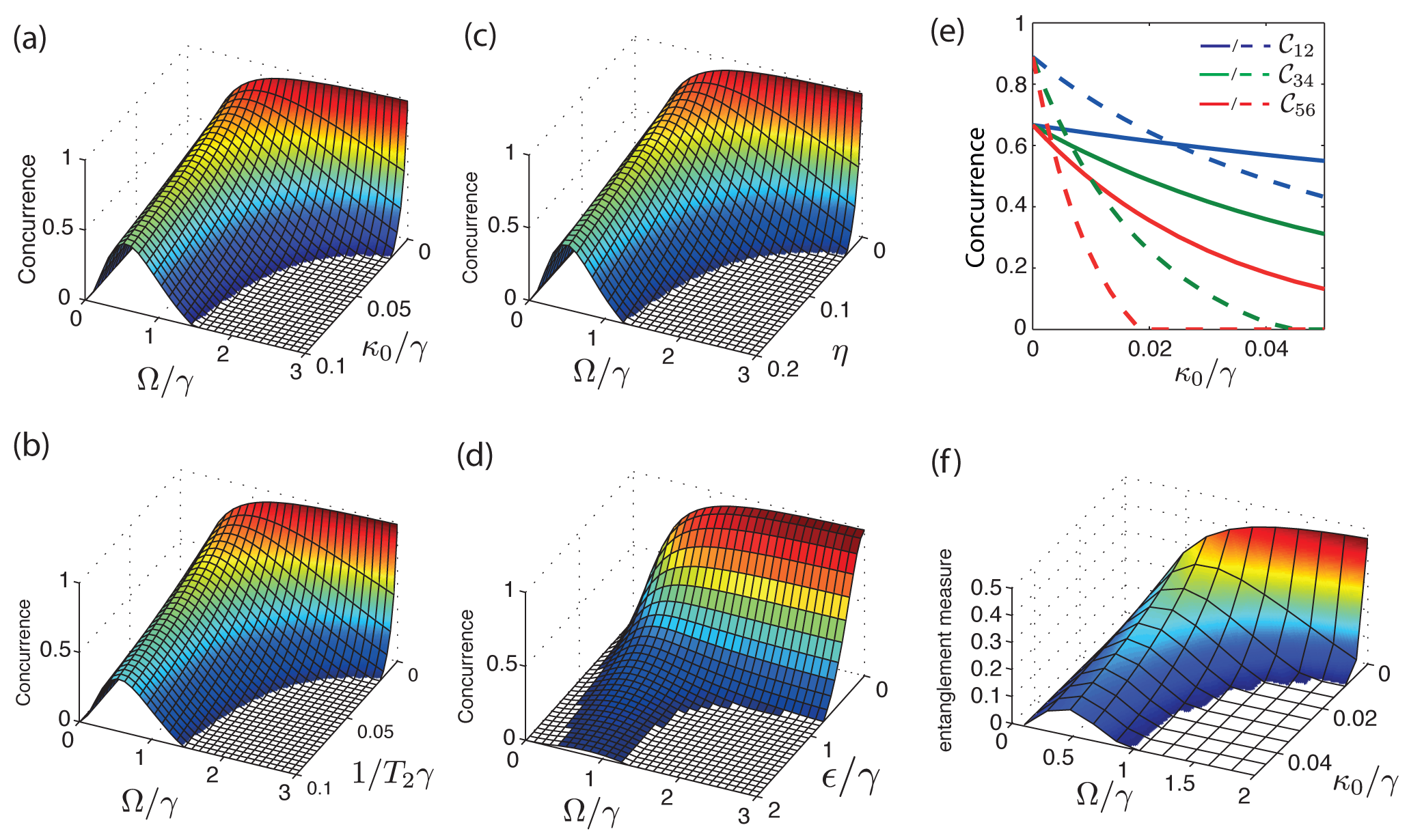}
\end{flushright}
\caption{Influence of imperfections on the entanglement properties of the steady state. (a-d) show the steady-state concurrence in a two-node network for various imperfections. (a) On-site decays (see text). 
(b) Dephasing of the spins modeled by additional Lindblad terms $\mL^\prime\rho=\sum_i\frac{1}{2 T_2}\mD[\sigma_z^i]\rho$.
(c) Waveguide losses, where $\eta$ is the fraction of photons that gets lost between the nodes (modeled by adding a factor $\sqrt{1-\eta}$ to the last term in \eeqref{eq:ME1}). (d) Deviation from the asymmetric detuning condition modeled by a symmetric  detuning offset $\delta_1=\delta_2=\epsilon$. In (a-c) we have chosen $\delta_1=\delta_2=0$.
(e) Influence of on-site decays on concurrences $\mathcal{C}_{i,i+1}$ of reduced two-spin density matrices $\rho_{i,i+1}$ for $N=6$ spins with $\delta_i=0$. 
(f) Influence of on-site decays on the steady state four-partite entanglement in a four-spin network, quantified by the measure of Ref.\,\cite{Jungnitsch2011}, which is bounded by $0.5$ in this case. The detuning profile is given by $\Delta^\prime=(0,-\gamma/2,\gamma/2,0)$. 
}
\label{fig:imperfections}
\end{figure}

\subsection{Imperfections}
\label{sec:imperfections}

Under realistic conditions various imperfections like onsite decays  or losses in the waveguide can violate the exact dark state condition and the system then evolves to a mixed (``bright") steady state. This is exemplified by the dashed lines in \fref{fig:brightAndDarkStates} for the case of onsite decays, modelled by adding an additional term $\mathcal{L}'\rho=\kappa_0 \sum_i \mathcal{D}[\sigma_-^i]\rho$ to the ME \eref{eq:ME2}. One clearly observes how the scattered intensity increases, while the purity drops as compared to the ideal case with $\kappa_0=0$.  
To study the entanglement properties in such non-ideal situations, we show in \fref{fig:imperfections}(a)-(d) the resulting steady state concurrence for various types of imperfections in a two-node network. In addition to onsite decays (panel (a)), we also consider intrinsic spin dephasing (panel (b)), waveguide losses (panel (c)), and small deviations from the ideal detuning profile (panel (d)). We see that different sources of imperfections lead to a qualitatively similar behavior and that in all cases the entanglement is quite robust and optimized for intermediate driving strengths $\Omega$.

For larger networks, the scattering of photons from the first nodes due to imperfections also affects successive spins, as shown in \fref{fig:imperfections}(e) for a six-spin dimer chain $\ket{\mS^0}$ in the presence of onsite losses. One clearly observes that the bi-partite entanglement in the dimers decreases with increasing number of previous nodes.
To study the robustness of genuine multi-partite entanglement in the presence of imperfections, we employ the entanglement measure proposed in Ref.~\cite{Jungnitsch2011}. It can be evaluated in a straight-forward way and \fref{fig:imperfections}(f) displays the results for a four-partite entangled steady state in the presence of onsite decays. The behavior of this measure qualitatively agrees with the results for the concurrence in the two-spin case (cf. \fref{fig:imperfections}(a)). 
That is, we observe a tradeoff between the maximal achievable entanglement and the robustness of the state. Finally, \fref{fig:brightAndDarkStates} also shows that for a fixed $N$, different bi- and multi-partite entangled states are affected equally.

\section{Cascaded non-linear cavities}
\label{sec:kerr}

%%%%%%%%%%%%%%%%%%%%%%%%%%%%%%%%%%%
%%%           FIGURE            %%%
%%%%%%%%%%%%%%%%%%%%%%%%%%%%%%%%%%%
\begin{figure}
\begin{flushright}
\includegraphics[width=0.8\textwidth]{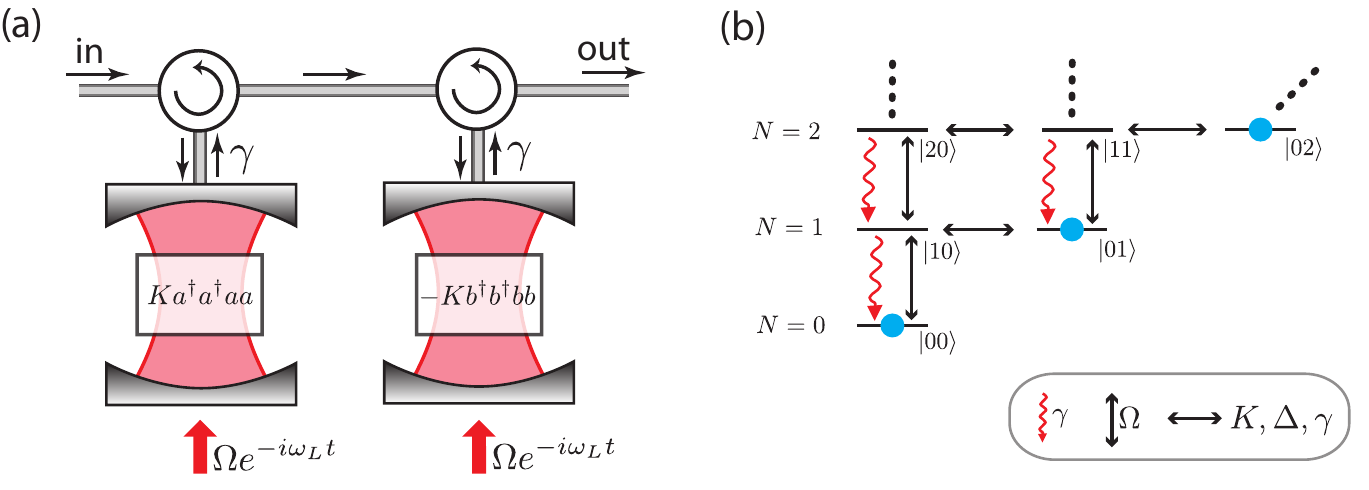}
\end{flushright}
\caption{(a) Two driven cascaded cavities with Kerr-type non-linearities. For simplicity, we assume the cavities to be driven through their back mirrors (red arrows), whereas the dominant decay channel $\gamma$ is provided by the front mirrors. (b) Level scheme for the construction of the pure steady state in the occupation number basis $\ket{nm}\equiv\ket{n}_+\ket{m}_-$ of the $c_\pm$ modes ($N=n+m$ is the total photon number). Straight arrows denote Hamiltonian matrix elements, while wavy ones correspond to the decay of the symmetric mode. Blue dots indicate population in the dark state \eqref{eq:kerr:chiState}. 
}
\label{fig:kerrSetup}
\end{figure}

The general concept of a coherent quantum absorber introduced in  \sref{sec:absorbers} suggests that the formation of dark entangled states can exist also for systems other than spins. As a second non-trivial example where this can be shown explicitly, we now discuss a setting where the two cascaded systems A and B are represented by two Kerr non-linear cavities as depicted in \fref{fig:kerrSetup}(a). The resulting distribution scheme for continuous variable entanglement is an alternative to other cascaded settings considered in this context \cite{Peng2002}.
We denote the two bosonic cavity modes by $a$ and $b$, and the dynamics of the system is governed by the ME \eqref{eq:ME2} with collective jump operator $c=a+b$ and cascaded Hamiltonian
\begin{eqnarray}
H_{\rm casc}=H_A+H_B -\rmi\frac{\gamma}{2} (b^\dagger a - a^\dagger b)\,.
\end{eqnarray}
Here, the Hamiltonian of the first cavity (system A, frequency $\om_{c,A}$) is 
\begin{eqnarray}
\label{eq:kerr:HA}
H_A&=\Delta a^\dag a + K a^\dag a^\dag a a +\rmi \Omega (a^\dag - a) \,,
\end{eqnarray}
where we have already moved to a frame rotating at the frequency $\om_d$ of the external driving field, such that $\Delta=\om_{c,A}-\om_d$ is the corresponding detuning of the cavity frequency and $\Omega$ the associated driving strength. Further, $K$ denotes the strength of the Kerr non-linearity. Motivated by our analysis of the cascaded spin system above,  we assume that the Hamiltonian for the second cavity is given by 
\begin{eqnarray}
\label{eq:kerr:HB}
H_B&=-\Delta b^\dag b - K b^\dag b^\dag b b +\rmi \Omega(b^\dag - b)\,.
\end{eqnarray}
Here we have chosen all constants to be identical to those used in $H_A$, such that the first two terms have opposite sign as compared to system A. As demonstrated below, this \emph{educated guess} for $H_B$ ensures  that the cascaded ME exhibits a dark state.

\subsection{Steady state solution}
\label{sec:kerr:solution}

In order to show that the ME \eqref{eq:ME2} indeed exhibits a pure steady state for the local Hamiltonians chosen above we exploit once more the conditions given in \eeqref{eq:conditions}. From condition ({\bf I}), i.e. $(a+b)\ket{\psi_0}=0$, it is clear that the steady state should contain zero quanta in the symmetric mode.\footnote{As shown in Ref.~\cite{Kraus2008} a pure stationary state could also exist if $|\psi_0\rangle$ is an eigenstate of the jump operator, i.e. $(a+b)\ket{\psi_0}=\lambda^\prime \ket{\psi_0}$ with $\lambda^\prime\in\mathbb{C}$. 
However, in the present example this would imply that the symmetric mode is in a coherent state of amplitude $\lambda^\prime$.
We do not expect this due to the non-linearity of the problem and it can indeed be shown that such an ansatz does not lead to a stationary solution unless  $\lambda^\prime=0$.
} 
Therefore, it is convenient to change to symmetric and anti-symmetric modes $c_\pm=(a\pm b)/\sqrt{2}$ and to write the dark state ansatz as $\ket{\psi_0}=\ket{0}_+ \ot \ket{\chi_0}_-$, with $\ket{\chi_0}=\sum_n\alpha_n\ket{n}$, where $\ket{n}$ are Fock states in the occupation number basis. In order to exploit condition ({\bf II}) we rewrite the cascaded Hamiltonian in terms of the symmetric and anti-symmetric modes,
\begin{eqnarray}
H_{\rm casc}=\rmi\sqrt{2}\Omega c_+^\dag  + \left(\Delta -\rmi\gamma/2 + K (\hat N-1) \right)\,c_+^\dag c_- + \hc \,.
\end{eqnarray}
Here $\hat N=c^\dag_+ c_+ + c^\dag_- c_-$ is the total number of quanta, which is conserved by all terms except for those $\propto\Omega$. 
Condition ({\bf II}) is equivalent to $H_{\rm casc}\ket{\psi_0}=\lambda\ket{\psi_0}$, and by projecting this equation onto $\ket{\psi_0}$ we see that it can only be fulfilled for $\lambda=0$ and hence 
\begin{eqnarray}
\label{eq:kerr:recursion}
\alpha_n=\sqrt{\frac{2}{n}} \frac{\epsilon}{x+n-1}\alpha_{n-1}\,,\quad
\epsilon=\frac{\Omega}{\rmi K}\,,\quad
x=\frac{i\Delta+\gamma/2}{\rmi K}\,.
\end{eqnarray}
This recursion is readily solved, such that the unique solution of conditions ({\bf I}) and ({\bf II}) is given by
\begin{eqnarray}
\label{eq:kerr:chiState}
\ket{\psi_0}=\ket{0}_+\otimes\ket{\chi_0}_-\,,\quad 
\textrm{with}\,\,\,
\ket{\chi_0}=\frac{1}{\mN}\sum_{n=0}^\infty 
\frac{(\sqrt{2}\epsilon)^n}{\sqrt{n!}} \frac{\Gamma(x)}{\Gamma(x+n)}
\ket{n}\,.
\end{eqnarray}
Here $\mN=\left[{}_0F_2(x,x^*;2\sabs{\epsilon}^2)\right]^{1/2}$ and ${}_0F_2$ denotes the generalized hyper-geometric function \cite{Gradshteyn2007}. Due to the simple tensor-product structure of $\ket{\psi_0}$ its only non-vanishing normally ordered moments are those of the $c_-$-mode, i.e.
\begin{eqnarray*}
\smean{(c_-^\dag)^n (c_-)^m}=
2^{\frac{n+m}{2}}
\frac{(\epsilon^*)^n\epsilon^m \Gamma(x^*)\Gamma(x)}{\Gamma(x^*+n)\Gamma(x+m)}
\frac{{}_0F_2(x^*+n,x+m;2\abs{\epsilon}^2)}{{}_0F_2(x^*,x;2\abs{\epsilon}^2)}\,,
\end{eqnarray*}
and the moments of the original modes thus read 
\begin{eqnarray}
\label{eq:kerr:moments}
\smean{(a^\dag)^n (b^\dag)^k (b)^l (a)^m}=
\frac{(-1)^{k+l}}{2^{(n+k+l+m)/2}} 
\smean{(c_-^\dag)^{n+k} (c_-)^{l+m}}\,.
\end{eqnarray}
Numerical studies suggest that there are no additional mixed steady states.

\subsection{Entanglement properties of the steady state}

%%%%%%%%%%%%%%%%%%%%%%%%%%%%%%%%%%%
%%%           FIGURE            %%%
%%%%%%%%%%%%%%%%%%%%%%%%%%%%%%%%%%%
\begin{figure}
\begin{flushright}
\includegraphics[width=0.8\textwidth]{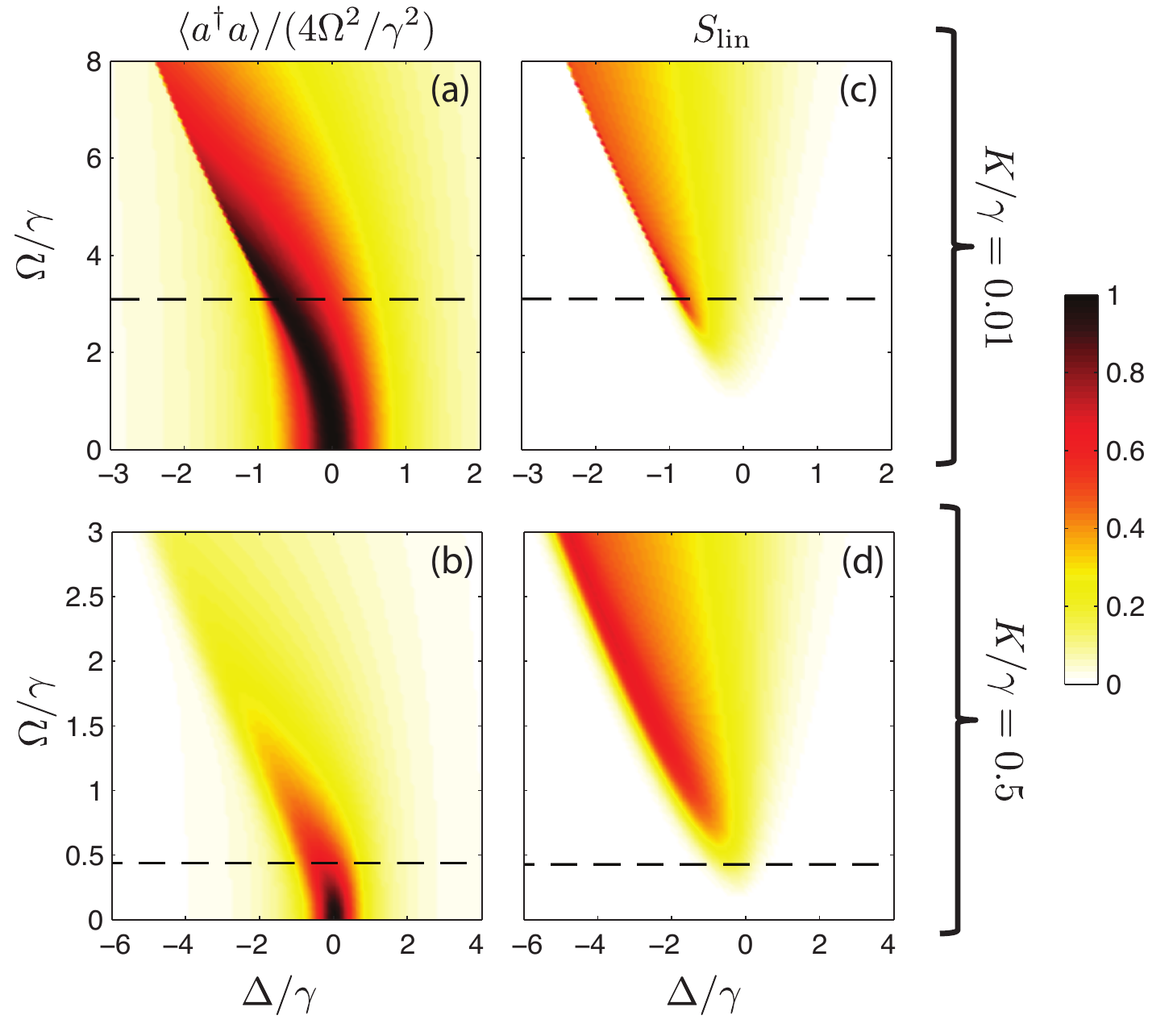}
\end{flushright}
\caption{Characterization of the steady state $\ket{\psi_0}$ for $K/\gamma=0.01$ (first row) and $K/\gamma=0.5$ (second row). (a),(b) photon number in the first cavity normalized to resonant linear response. (c),(d) linear entropy $S_{\rm lin}$ of the first cavity as defined in \eeqref{eq:kerr:Slin}. 
In all panels the dashed lines indicate the critical driving strength $\Omega_c$, above which the semi-classical response becomes bi-stable (see, e.g., Ref.\,\cite{Yurke2006}).}
\label{fig:kerrFigure}
\end{figure}

Despite the simple structure of the steady state $\ket{\psi_0}$ in the $c_\pm$ representation, it is generally entangled when written in terms of the  original modes $a$ and $b$ according to
\begin{eqnarray}
\label{eq:kerr:B}
\ket{\psi_0}=\sum_{n,m=0}^\infty B_{nm} \ket{n}_A\ket{m}_B\,,
\,\,
B_{nm}=\alpha_{n+m}\, \frac{(-1)^m}{2^{(n+m)/2}}
\sqrt{\frac{(n+m)!}{n! m!}},
\end{eqnarray}
where the $\alpha_n$ can be read off from \eeqref{eq:kerr:chiState}.
We characterize the entanglement with respect to this bipartition by the mixedness of the reduced state $\rho_A^0=\tr_B\{\pr{\psi_0}\}=BB^\dag$ of the first cavity, as measured by its linear entropy
\begin{eqnarray}
\label{eq:kerr:Slin}
S_{\rm lin}=1-\tr\{(\rho^0_A)^2\}\,.
\end{eqnarray}
The results for two different strengths of the non-linearity are displayed in \fref{fig:kerrFigure}. For a better orientation, we show in panels (a) and (b) the photon number of the first cavity, normalized to the resonant response in the linear case, i.e. $\Delta=K=0$. Here, one clearly observes the well-known behavior of a single Kerr-non-linear cavity, which is characterized by a deformation of the response curve for increasing $\Omega$. Above a certain critical driving strength $\Om_c$ the classical response becomes bi-stable, while the quantum mechanical response curve exhibits a sharp step \cite{Drummond1980}. From panels (c) and (d) we see that the regions of pronounced non-linear response are those where the entropy is large, signaling a high degree of entanglement between the two cavities. In general, the features of the response are more washed out for larger non-linearities, where also the regions of high entropy are more extended. Note that the state $\ket{\chi_0}$, and hence also $\ket{\psi_0}$, is generally non-Gaussian, as can also be shown by evaluating appropriate measures \cite{Geroni2007}. Finally, we remark that for $K\rightarrow 0$ the steady state approaches a product of coherent states, i.e. $\ket{\psi_0}\rightarrow\ket{\beta}_A\otimes\ket{-\beta}_B$, with  $\beta=\epsilon/x$, where $\ket{\beta}$ denotes a coherent state of amplitude $\beta$. We are thus able to recover the classical limit in which no entanglement persists between the cavities.

\subsection{The inverse coherent absorber problem}  

Finally, let us take a different view on the analysis presented so far and emphasize once more that cascaded systems may be solved in a successive fashion ``from left to right". This means that tracing the full cascaded ME \eqref{eq:ME2} over the second subsystem B yields a closed equation for the first subsystem A, which in this case reads
\begin{eqnarray}
\label{eq:kerr:MEA}
\dot\rho_A=\mL_A\rho_A \equiv -\rmi[\Delta a^\dag a + K a^\dag a^\dag a a +\rmi \Omega (a^\dag - a),\rho_A] + \gamma\mD[a]\rho_A\,.
\end{eqnarray}
In  \sref{sec:absorbers} we have assumed that the stationary \emph{mixed} state solution of this equation including its spectral decomposition is known, which allowed us to explicitly construct a corresponding perfect absorber system B, such that the whole system is driven into a pure steady state. 
However, in many cases solving for the steady state $\rho_A^0$ of \eeqref{eq:kerr:MEA} is a non-trivial problem by itself, which in the present case was first accomplished using phase space methods \cite{Drummond1980}.
Note that by using an educated guess for the Hamiltonian $H_B$  of the absorber system and then solving for the \emph{pure} steady state $\ket{\psi_0}$ of the whole network, we have also indirectly solved the mixed state problem of a single cavity by computing $\rho_A^0=\tr_B\{\pr{\psi_0}\}$. Its explicit matrix-elements in the Fock-state basis agree with the expressions found in the literature \cite{Kheruntsyan1999}, as do its moments $\smean{(a^\dag)^n (a)^m}$ obtained as a special case of \eeqref{eq:kerr:moments} \cite{Drummond1980}. The procedure presented above thus represents an elegant alternative way of obtaining the mixed steady state of the dispersive optical bi-stability problem.Ê

Avoiding to work with mixed states by using pure states on larger Hilbert spaces has found wide-spread use in the quantum information community, one of the reasons being that pure states can be manipulated more easily \cite{Nielsen2000}. From this point of view, the steady state $\ket{\psi_0}$ possesses additional relevance as a \emph{purification} of the density matrix $\rho_A^0$. This is an interesting result in itself, since obtaining such a purification generally requires knowledge of the spectral decomposition of $\rho_A^0$---a strong requirement even if $\rho_A^0$ is known explicitly. In the above example we were able to circumvent this difficulty by directly constructing the purification $\ket{\psi_0}$ as a steady state of the cascaded ME. Although so far this procedure is limited to systems where the corresponding absorber Hamiltonian $H_B$ can be obtained by an \emph{educated guess},  it is intriguing to think about cascaded quantum systems also as an analytic tool for calculating stationary states of non-trivial open quantum systems.

\section{Implementations}
\label{sec:implementations}

The two key ingredients for realizing cascaded networks as discussed in the previous sections are
(i) the implementation of low-loss non-reciprocal devices for directional routing of photons, and
(ii) achieving a coupling of single quantum systems to a 1D waveguide that exceeds local decoherence channels. In the following, we discuss several implementations that fulfill these requirements, where a focus lies on optical and microwave setups.
Requirement (i) is crucial for realizing the unidirectional coupling between the nodes and can in principle be fulfilled by standard circulators based on the Faraday effect (see, e.g., Ref.\,\cite{Gripp1995} for an optical implementation), but also by exploiting unidirectional edge modes in media with broken time-reversal symmetry \cite{Haldane2008,Wang2008,Wang2009}. However, in recent years there has been increasing interest in designing non-reciprocal on-chip devices which are integrable with, e.g., microwave circuitry \cite{Koch2010,Kamal2011} or nano-fabricated photonic components \cite{Wang2005,Feng2011}.
Such elements would allow one to build up cascaded networks in a very controlled way in both the optical and the microwave domain, and thus constitute a promising way of implementing the scenarios discussed in this work.

%%%%%%%%%%%%%%%%%%%%%%%%%%%%%%%%%%%
%%%           FIGURE            %%%
%%%%%%%%%%%%%%%%%%%%%%%%%%%%%%%%%%%
\begin{figure}
\begin{flushright}
\includegraphics[width=0.8\textwidth]{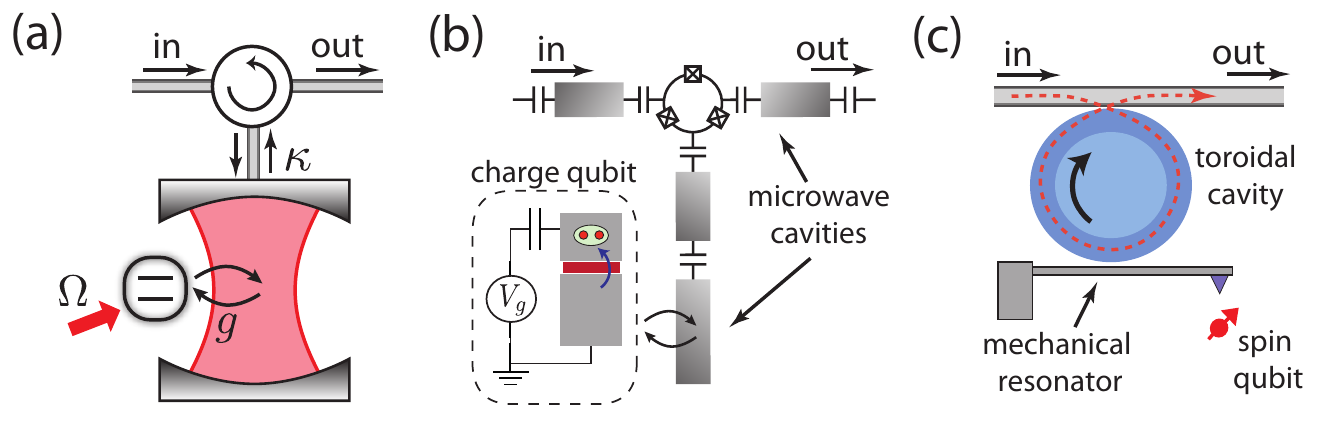}
\end{flushright}
\caption{Possible realizations of a single node of a cascaded spin network as shown in \fref{fig:figure1}. (a) generic setup where the unidirectional coupling is achieved by a cavity connected to a circulator, (b) non-reciprocal superconducting circuit  based on the proposal of Ref.~\cite{Koch2010}, (c) optomechanical transducer based on toroidal cavities \cite{Stannigel2010,Anetsberger2009}.}
\label{fig:implementations}
\end{figure}

Concerning the second requirement, we first discuss the case of the cascaded spin network presented in \sref{sec:spin}. Realizing a coupling of TLSs to a 1D waveguide could, e.g., be achieved along the lines of the experiments reported in Refs.\,\cite{Astafiev2010,Akimov2007}, where superconducting qubits or atoms are coupled to transmission lines or hollow core optical fibers, respectively. However, one can also do without a direct coupling of the TLS to the waveguide by using the generic and flexible approach depicted in \fref{fig:implementations}(a). Here, the TLS is coupled to a cavity, whose output port is connected to a circulator or another non-reciprocal device. In the bad cavity limit $\kappa\gg g$, where $g$ is the TLS-cavity coupling, a series of these nodes results in the desired model \eqref{eq:ME1}, with an effective TLS-waveguide decay rate $\gamma=2g^2/\kappa$. A particular realization of this scheme in the context of circuit cavity quantum electrodynamics  \cite{SchoelkopfNature2008,Koch2010} is shown in \fref{fig:implementations}(b). Finally note that other systems like optomechanical transducers (\fref{fig:implementations}(c))
have been proposed to realize a similar unidirectional coupling \cite{Stannigel2010,Anetsberger2009}.
In the case of cascaded  cavities discussed in \sref{sec:kerr}, the most important requirement is the realization of strong Kerr-type non-linearities, which are of equal magnitude and opposite sign in the two cavities. 
Such a tunable non-linearity can be realized by a dispersive coupling of the cavity field to a suitable four-level system, which has been analyzed both for optical~\cite{Imamoglu1997,Hartmann2008} as well as superconducting microwave cavities \cite{Rebic2009}.

In summary, we find that the current experimental capabilities for realizing strong TLS-photon interactions in various systems, combined with the development of novel non-reciprocal devices in the optical and microwave domain, enable the implementation and design of various cascaded quantum networks. Here, the dissipative state preparation schemes described in this work could serve as an interesting application exploring the unconventional physical properties of such devices.

\section{Conclusions} 
\label{sec:conclusion}

We have shown that photon emission and coherent reabsorption processes in cascaded quantum systems can lead to the formation of pure and highly entangled steady states. In the case of spin networks, this mechanism provides a tunable dissipative preparation scheme for a whole class of  multi-partite entangled states, and instances of this scheme might serve as a basis for dissipative quantum communication protocols \cite{Vollbrecht2011}. For the case of  two cascaded non-linear cavities we have identified a dissipative preparation scheme for non-Gaussian entangled states, and we have illustrated how cascaded systems can serve as a new analytic tool to evaluate the stationary states of driven-dissipative systems.
More generally, our findings show that such driven cascaded networks realize a novel type of non-equilibrium quantum many-body system, which can be implemented with currently developed integrated optical systems or superconducting devices.

\ack 

The authors thank B. Kraus for valuable discussions.
This work was supported by the EU network AQUTE and the Austrian Science Fund (FWF) through SFB FOQUS and the START grant Y 591-N16. 

\appendix

\section{General construction of coherent quantum absorbers}
\label{sec:app:absorber}

We present the arguments leading to the results quoted in \sref{sec:constructionAbsorbers}. Given a system A in terms of its Hamiltonian $H_A$ and jump operator $c_A$, we seek a suitable system B described by $H_B$ and $c_B$ which perfectly reabsorbs the output field of A in steady state. In the following, we construct system B by requiring that the total cascaded system evolves into a pure steady state. The dynamics of the network is governed by the ME \eref{eq:ME2} with Hamiltonian
\begin{equation}
H_{\rm casc}= H_A+H_B- \rmi \frac{\gamma}{2}(c_A c_B^\dag  - c^\dag_A c_B )\,
\end{equation} 
and  collective jump operator $c=c_A+c_B$, and assuming that $c$ has no eigenvectors a pure state $\ket{\psi_0}$ is a stationary state of \eeqref{eq:ME2} if and only if \cite{Kraus2008}
\begin{eqnarray}
c\ket{\psi_0}=0\,,\qquad
[H_{\rm casc},\pr{\psi_0}]=0\,.
\end{eqnarray}
As discussed in \sref{sec:constructionAbsorbers}, the cascaded nature of the interaction allows us to solve for the reduced steady state $\rho_A^0$ of system A without knowing anything about B. Assuming uniqueness, we introduce its spectral decomposition 
\begin{eqnarray}
\rho_A^0=\sum_k p_k \pr{k}\,,
\end{eqnarray}
and assume that the eigenvalues $p_k$ are positive and non-degenerate.
A potential pure steady state $\ket{\psi_0}$  of the whole system can then be written as a purification of $\rho_A^0$, i.e.,  
$|\psi_0\rangle=\sum_k \sqrt{p_k} |k\rangle_A\otimes |\tilde k\rangle_B$ \cite{Nielsen2000}. Here,  $|\tilde k\rangle= V|k\rangle$, and $V$ denotes an arbitrary unitary and we assume subsystem B to have the same Hilbert space dimension as A. 

In the following, we write out the tensor-products more explicitly, such that the dark-state condition $c\ket{\psi_0}=0$ reads
\begin{eqnarray}
\left(c_A\ot\id + \id\ot c_B\right)\ket{\psi_0}=0\,,
\label{eq:app:darkStateCond}
\end{eqnarray}
with 
\begin{eqnarray*}
c_A&=\sum_{n,m}   \langle n | c_A|m\rangle  |n\rangle\langle m|\,, \quad
c_B=\sum_{n,m}   \langle \tilde n | c_B|\tilde m\rangle  |\tilde n\rangle\langle \tilde m|\,. 
\end{eqnarray*}
To proceed, we plug the ansatz for $|\psi_0\rangle$ into \eeqref{eq:app:darkStateCond},
yielding
\begin{equation*}
%\begin{split}
\sum_{k, n} \sqrt{p_k} \left(   \langle n | c_A|k\rangle |n\rangle\otimes | \tilde k\rangle + \langle \tilde n | c_B|\tilde k\rangle |k\rangle\otimes | \tilde n\rangle\right)=0\,,
%\end{split}
\end{equation*}
and after relabeling  the indices we obtain
\begin{equation}
%\begin{split}
\sum_{k, n}   \left(  \sqrt{p_k} \langle n | c_A|k\rangle  +   \sqrt{p_n}\langle \tilde k | c_B|\tilde n\rangle \right) |k\rangle\otimes | \tilde n\rangle=0\,.
%\end{split}
\end{equation}
Since the $\ket{k}$ form an orthonormal basis, this condition is equivalent to
\begin{equation}
\label{eq:relationcAcB}
\sqrt{p_n} \langle \tilde k | c_B|\tilde n\rangle= -  \sqrt{p_k} \langle n | c_A|k\rangle\,,
\end{equation}
which fixes the operator $c_B$ to be
\begin{eqnarray}
\label{eq:app:cB}
c_B&=& - \sum_{n,m}   \sqrt{\frac{p_n}{p_m}}   \langle m| c_A| n\rangle  |\tilde n\rangle\langle \tilde m|\,.
\end{eqnarray}

To construct the Hamiltonian $H_B$ for system B we exploit the second condition  $[H_{\rm casc},\pr{\psi_0}]=0$ needed for a pure steady state, which is equivalent to $H_{\rm casc}\ket{\psi_0}=\lambda\ket{\psi_0}$ for $\lambda\in\mathbbm{R}$. Note that $c|\psi_0\rangle=0$ implies 
\begin{equation*}
\rmi \frac{\gamma}{2}(c_A\ot c_B^\dag  - c^\dag_A \ot c_B ) |\psi_0\rangle 
=  \rmi \frac{\gamma}{2}(c^\dag_A c_A \ot  \id  -   \id \ot  c^\dag_B c_B ) |\psi_0\rangle\,,
\end{equation*} 
such that this  condition reads 
\begin{equation}
\label{eq:app:stationarityCond}
\left(H_{A,\eff}\ot\id+\id\ot H^\dag_{B,\eff}\right)|\psi_0\rangle = \lambda |\psi_0\rangle\,,
\end{equation}
where we have introduced the effective non-hermitian Hamiltonians $H_{j,{\rm eff}}=H_j-\rmi\frac{\gamma}{2}c_j^\dag c_j$. Since a finite $\lambda$ would only lead to a global shift of $H_B$ below, we can assume $\lambda=0$ without loss of generality. We write the Hamiltonian as 
\begin{eqnarray}
H_B&=\sum_{n,m}   \langle \tilde n | H_B|\tilde m\rangle  |\tilde n\rangle\langle \tilde m|\,,
\end{eqnarray}
and to determine the matrix elements in this expansion we start from \eeqref{eq:app:stationarityCond} and proceed as for the dark state condition \eeqref{eq:app:darkStateCond} with the replacements $c_A\rightarrow H_{A,\eff}$ and $c_B\rightarrow H_{B,\eff}^\dag$. As an intermediate result this yields
\begin{equation}
%\begin{split}
\label{eq:app:HBintermediate}
\langle \tilde n | H_{B} |\tilde m\rangle= -  \sqrt{\frac{p_n}{p_m}} \langle m | H_{A,{\rm eff}}|n\rangle-\rmi\frac{\gamma}{2} \langle \tilde n | c^\dag_{B}c_B |\tilde m\rangle \,.
%\end{split}
\end{equation}
To express the right-hand side of this equation fully in terms of operators on A, we employ the identity
\begin{equation}
\langle \tilde n | c^\dag_{B}c_B |\tilde m\rangle= \sqrt{\frac{1}{p_m p_n}} \langle m | c_A\rho_A^0 c_A^\dag| n\rangle\,,
\end{equation}
which can be derived with the help of \eeqref{eq:relationcAcB},
and then make use of the stationarity of system A,
\begin{equation}
\gamma c_A\rho_A^0 c_A^\dag= \rmi (H_{A,{\rm eff}} \rho_A^0 - \rho_A^0 H_{A,{\rm eff}}^\dag )\,.
\end{equation}
\Eref{eq:app:HBintermediate} then becomes
\begin{eqnarray}
\label{eq:app:HB}
\langle \tilde n | H_{B} |\tilde m\rangle
= -  \frac{1}{2}\left( \sqrt{\frac{p_n}{p_m}} \langle m | H_{A,{\rm eff}}|n\rangle  + \sqrt{\frac{p_m}{p_n}} \langle m | H_{A,{\rm eff}}^\dag |n\rangle \right) \,,
\end{eqnarray}
which determines the Hamiltonian of system B. The expressions \eqref{eq:app:cB} and \eqref{eq:app:HB} are the results quoted in \sref{sec:constructionAbsorbers} of the main text.

\section{Dark states of cascaded spin networks}
\label{sec:app:spin}

We provide details regarding the cascaded spin network presented in \sref{sec:spin}. The system is described by the ME \eqref{eq:ME2} with  collective jump operator $c=\sum_i\sigma^i_-$ and $H_{\rm casc}$ defined in \eeqref{eq:Hcasc}.

\subsection{Uniqueness of the steady state $\ket{\stState^0}$}

We show by explicit construction that the state $\ket{\stState^0}$ given in \eeqref{eq:psi0} is the unique steady state of the network, provided that $\delta_{2i-1}=-\delta_{2i}$ and $\Omega_i=\Omega$ for $i=1,2,\ldots$. To do so, we exploit the fact that the cascaded interaction allows for a successive construction of the steady state and start with the case $N=2$. In this case, the steady state $\rho_2^0$ is obtained by solving $\mL^{(2)}\rho_2^0=0$, where  $\mL^{(2)}=\mL_{12}$ is given by the block-wise Liouvillian
\begin{eqnarray}
\mL_{i,i+1}\rho=&-\rmi\left[\frac{\delta_i}{2}(\sigma^i_z-\sigma^{i+1}_z) + \Omega(\sigma^i_x + \sigma^{i+1}_x)\,,\,\rho\right] +\frac{\gamma}{2}\mD[c_{i,i+1}]\rho
\end{eqnarray}
with $c_{i,j}=\sigma^i_-+\ldots +\sigma^j_-$. We have already seen in the main text that a solution is given by $\rho_2^0=\pr{\stState_{\delta_1}}$, and its uniqueness can be shown by calculating the characteristic polynomial of $\mL_{12}$ and realizing that there is only one zero eigenvalue for $\gamma>0$.

We continue with $N=4$ and write the ansatz for the steady state as $\rho_4^0=\pr{\stState_{\delta_1}}\otimes \mu$, where $\mu$ is a two-node density matrix. The four-node Liouvillian can be rewritten as
\begin{eqnarray}
\mL^{(4)}\rho= \mL_{12}\rho + \mL_{34}\rho 
-\gamma \left( [c^\dag_{34},c_{12}\rho] + [\rho c^\dag_{12},c_{34}] \right)\,,
\end{eqnarray}
and we note  that $\mL_{12}\pr{\stState_{\delta_1}}=0$ as well as $c_{12}\ket{\stState_{\delta_1}}=0$, such that the equation $\mL^{(4)}\rho_4^0=0$ simplifies to $\mL_{34}\mu=0$. However, this is just the two-node problem we have already solved and the unique solution for the four-node netwrok is thus given by $\rho_4^0=\pr{\stState_{\delta_1}}\otimes\pr{\stState_{\delta_3}}$. By iterating this argument in blocks of two spins, we obtain the steady state $\ket{\stState^0}$ given in \eeqref{eq:psi0}.

\subsection{Unitary form invariance of the master equation}
\label{app:formInvariance}

We briefly demonstrate that the cascaded Hamiltonian of the spin network is form-invariant under the unitary transformations \eref{eq:U} as stated in \eeqref{eq:formInvariance}. In order to calculate $U_i H_{\rm casc} U_i^\dag$, we write $j\equiv i+1$ for brevity and rearrange $H_{\rm casc}$ as follows:
\begin{eqnarray*}
H_{\rm casc}&=\sum_{k\neq i,j}\left( \frac{\delta_k}{2}\sigma_z^k+ \Omega \sigma_x^k \right)
-\rmi\frac{\gamma}{2}\sum_{k>l}^{k,l\neq i,j}(\sigma_+^k\sigma_-^l - \sigma_-^k\sigma_+^l)\\
&-\rmi\frac{\gamma}{2}\left( c^\dagger_{i,j}c_{1,i-1} + c^\dagger_{j+1,N} c_{i,j} -\hc  \right)\\
&+\frac{\delta_i+\delta_j}{2} \frac{1}{2}(\sigma_z^i+\sigma_z^j)
+\Omega(\sigma_x^i+\sigma_x^j)\\
\label{eq:lastLineTransf}
&+\left[
\frac{\delta_i-\delta_j}{2} \frac{1}{2}(\sigma_z^i-\sigma_z^j)
-\rmi\frac{\gamma}{2}(\sigma_+^j \sigma_-^i - \sigma_-^j \sigma_+^i)
\right] \,,
\end{eqnarray*}
where we have again used the piecewise jump operator $c_{k,l}=\sigma^k_-+\ldots +\sigma^l_-$. Note that $U_i(\theta)$ commutes with the first three lines and we thus focus on the last one. We abbreviate the operators appearing there by $A=(\sigma_z^i-\sigma_z^j)/2$ and $B=\sigma^j_+\sigma^i_- - \sigma^j_- \sigma^i_+$ and observe that they transform into one another:
\begin{eqnarray}
U_{i}(\theta) A U_{i}^\dag(\theta)= A \cos 2\theta + i B \sin 2\theta\\
U_{i}(\theta) B U_{i}^\dag(\theta)= B \cos 2\theta + i A \sin 2\theta
\end{eqnarray}
Introducing the difference of detunings $\delta=(\delta_i-\delta_j)/2$, a non-trivial form-invariance of the Hamiltonian $H_{\rm casc}$ under $U_i(\theta)$ is realized if 
\begin{eqnarray}
& U_{i}(\theta)\left[ \delta A -\rmi \frac{\gamma}{2} B \right] U_{i}^\dag(\theta)
=-\delta A -\rmi \frac{\gamma}{2} B
\end{eqnarray}
That is, we require the detunings to swap and the cascaded part to remain invariant. It is easy to check that this requirement results in two equations, which are both solved for the choice $\tan\theta=-2\delta/\gamma=(\delta_{i+1}-\delta_i)/\gamma$, which was to be demonstrated.

\section*{Bibliography}

\bibliographystyle{unsrt}

\bibliography{CascadedQSexport}

\begin{thebibliography}{10}

\bibitem{Nielsen2000}
M.~A. Nielsen and I.~L. Chuang.
\newblock {\em Quantum computation and quantum information}.
\newblock Cambridge University Press, Cambridge, 2000.

\bibitem{Metcalf1999}
H.~J. Metcalf and P.~van Straten.
\newblock {\em Laser cooling and trapping}.
\newblock Graduate texts in contemporary physics. Springer, New York, 1999.

\bibitem{Wiseman2010}
Howard~M. Wiseman and Gerard~J. Milburn.
\newblock {\em Quantum measurement and control}.
\newblock Cambridge University Press, Cambridge, 1. publ. edition, 2010.

\bibitem{Mueller2012}
M.~M{\"u}ller, S.~Diehl, G.~Pupillo, and P.~Zoller.
\newblock Engineered open systems and quantum simulations with atoms and ions.
\newblock arXiv:1203.6595, 2012.

\bibitem{Plenio1999}
M.~B. Plenio, S.~F. Huelga, A.~Beige, and P.~L. Knight.
\newblock Cavity-loss-induced generation of entangled atoms.
\newblock {\em Phys. Rev. A}, 59:2468, Mar 1999.

\bibitem{Parkins2000}
A.~S. Parkins and H.~J. Kimble.
\newblock Position-momentum {Einstein-Podolsky-Rosen} state of distantly
  separated trapped atoms.
\newblock {\em Phys. Rev. A}, 61:052104, Apr 2000.

\bibitem{Clark2003}
S.~Clark, A.~Peng, M.~Gu, and S.~Parkins.
\newblock Unconditional preparation of entanglement between atoms in cascaded
  optical cavities.
\newblock {\em Phys. Rev. Lett.}, 91:177901, Oct 2003.

\bibitem{Kraus2004}
B.~Kraus and J.~I. Cirac.
\newblock Discrete entanglement distribution with squeezed light.
\newblock {\em Phys. Rev. Lett.}, 92:013602, Jan 2004.

\bibitem{Paternostro2004}
M.~Paternostro, W.~Son, and M.~S. Kim.
\newblock Complete conditions for entanglement transfer.
\newblock {\em Phys. Rev. Lett.}, 92:197901, May 2004.

\bibitem{Parkins2006}
A.~S. Parkins, E.~Solano, and J.~I. Cirac.
\newblock Unconditional two-mode squeezing of separated atomic ensembles.
\newblock {\em Phys. Rev. Lett.}, 96:053602, Feb 2006.

\bibitem{Kastoryano2011}
M.~J. Kastoryano, F.~Reiter, and A.~S. S\o{}rensen.
\newblock Dissipative preparation of entanglement in optical cavities.
\newblock {\em Phys. Rev. Lett.}, 106:090502, Feb 2011.

\bibitem{Diehl2008}
S.~Diehl, A.~Micheli, A.~Kantian, B.~Kraus, H.~P. B\"uchler, and P.~Zoller.
\newblock {Quantum states and phases in driven open quantum systems with cold
  atoms}.
\newblock {\em Nat Phys}, 4:878, November 2008.

\bibitem{Kraus2008}
B.~Kraus, H.~P. B\"uchler, S.~Diehl, A.~Kantian, A.~Micheli, and P.~Zoller.
\newblock Preparation of entangled states by quantum markov processes.
\newblock {\em Phys. Rev. A}, 78:042307, Oct 2008.

\bibitem{Verstraete2009}
F.~Verstraete, M.~M. Wolf, and J.~I. Cirac.
\newblock {Quantum computation and quantum-state engineering driven by
  dissipation}.
\newblock {\em Nat Phys}, 5:633, 2009.

\bibitem{Weimer2010}
H.~Weimer, M.~M{\"u}ller, I.~Lesanovsky, P.~Zoller, and H.~P. B{\"u}chler.
\newblock A rydberg quantum simulator.
\newblock {\em Nat Phys}, 6:382, 2010.

\bibitem{Cho2011}
J.~Cho, S.~Bose, and M.~S. Kim.
\newblock Optical pumping into many-body entanglement.
\newblock {\em Phys. Rev. Lett.}, 106:020504, Jan 2011.

\bibitem{Vollbrecht2011}
K.~G.~H. Vollbrecht, C.~A. Muschik, and J.~I. Cirac.
\newblock Entanglement distillation by dissipation and continuous quantum
  repeaters.
\newblock {\em Phys. Rev. Lett.}, 107:120502, Sep 2011.

\bibitem{Pastawski2011}
F.~Pastawski, L.~Clemente, and J.~I. Cirac.
\newblock Quantum memories based on engineered dissipation.
\newblock {\em Phys. Rev. A}, 83:012304, Jan 2011.

\bibitem{Barreiro2011}
J.~T. Barreiro, M.~M{\"u}ller, P.~Schindler, D.~Nigg, T.~Monz, M.~Chwalla,
  M.~Hennrich, C.~F. Roos, P.~Zoller, and R.~Blatt.
\newblock An open-system quantum simulator with trapped ions.
\newblock {\em Nature}, 470:486, 2011.

\bibitem{Krauter2011}
H.~Krauter, C.~A. Muschik, K.~Jensen, W.~Wasilewski, J.~M. Petersen, J.~I.
  Cirac, and E.~S. Polzik.
\newblock Entanglement generated by dissipation and steady state entanglement
  of two macroscopic objects.
\newblock {\em Phys. Rev. Lett.}, 107:080503, Aug 2011.

\bibitem{Muschik2011}
C.~A. Muschik, E.~S. Polzik, and J.~I. Cirac.
\newblock Dissipatively driven entanglement of two macroscopic atomic
  ensembles.
\newblock {\em Phys. Rev. A}, 83:052312, May 2011.

\bibitem{Gardiner1993}
C.~W. Gardiner.
\newblock Driving a quantum system with the output field from another driven
  quantum system.
\newblock {\em Phys. Rev. Lett.}, 70:2269, Apr 1993.

\bibitem{Carmichael1993}
H.~J. Carmichael.
\newblock Quantum trajectory theory for cascaded open systems.
\newblock {\em Phys. Rev. Lett.}, 70:2273, Apr 1993.

\bibitem{Kane1997}
C.~L. Kane and M.~P.~A. Fisher.
\newblock In S.~Das Sarma and A.~Pinczuk, editors, {\em Perspectives in the
  Quantum Hall effect}. Wiley, 1997.

\bibitem{Haldane2008}
F.~D.~M. Haldane and S.~Raghu.
\newblock Possible realization of directional optical waveguides in photonic
  crystals with broken time-reversal symmetry.
\newblock {\em Phys. Rev. Lett.}, 100:013904, Jan 2008.

\bibitem{Wang2008}
Z.~Wang, Y.~Chong, J.~D. Joannopoulos, and M.~Solja\ifmmode \check{c}\else
  \v{c}\fi{}i\ifmmode~\acute{c}\else \'{c}\fi{}.
\newblock Reflection-free one-way edge modes in a gyromagnetic photonic
  crystal.
\newblock {\em Phys. Rev. Lett.}, 100:013905, Jan 2008.

\bibitem{Wang2009}
Z.~Wang, Y.~Chong, J.~D. Joannopoulos, and M.~Solja\ifmmode \check{c}\else
  \v{c}\fi{}i\ifmmode~\acute{c}\else \'{c}\fi{}.
\newblock Observation of unidirectional backscattering-immune topological
  electromagnetic states.
\newblock {\em Nature}, 461:772, 10 2009.

\bibitem{Hafezi2011}
M.~Hafezi, E.~A. Demler, M.~D. Lukin, and J.~M. Taylor.
\newblock Robust optical delay lines with topological protection.
\newblock {\em Nat. Phys.}, 7:907, 2011.

\bibitem{Wang2005}
Z.~Wang and S.~Fan.
\newblock Optical circulators in two-dimensional magneto-optical photonic
  crystals.
\newblock {\em Opt. Lett.}, 30:1989, Aug 2005.

\bibitem{Feng2011}
L.~Feng, M.~Ayache, J.~Huang, Y.-L. Xu, M.-H. Lu, Y.-F. Chen, Y.~Fainman, and
  A.~Scherer.
\newblock Nonreciprocal light propagation in a silicon photonic circuit.
\newblock {\em Science}, 333:729, 2011.

\bibitem{Koch2010}
J.~Koch, A.~A. Houck, K.~Le~Hur, and S.~M. Girvin.
\newblock Time-reversal-symmetry breaking in circuit-qed-based photon lattices.
\newblock {\em Phys. Rev. A}, 82:043811, Oct 2010.

\bibitem{Kamal2011}
A.~Kamal, J.~Clarke, and M.~H. Devoret.
\newblock Noiseless non-reciprocity in a parametric active device.
\newblock {\em Nat Phys}, 7:311, Apr 2011.

\bibitem{Drummond1980}
P.~D. Drummond and D.~F. Walls.
\newblock {Quantum theory of optical bistability. I. Nonlinear polarisability
  model}.
\newblock {\em Journal of Physics A: Mathematical and General}, 13:725, 1980.

\bibitem{Stannigel2011}
K.~Stannigel, P.~Rabl, A.~S. S\o{}rensen, M.~D. Lukin, and P.~Zoller.
\newblock Optomechanical transducers for quantum-information processing.
\newblock {\em Phys. Rev. A}, 84:042341, Oct 2011.

\bibitem{Gu2006}
Mile Gu, A.~S. Parkins, and H.~J. Carmichael.
\newblock Entangled-state cycles from conditional quantum evolution.
\newblock {\em Phys. Rev. A}, 73:043813, Apr 2006.

\bibitem{Jungnitsch2011}
B.~Jungnitsch, T.~Moroder, and O.~G\"uhne.
\newblock Taming multiparticle entanglement.
\newblock {\em Phys. Rev. Lett.}, 106:190502, May 2011.

\bibitem{Peng2002}
A.~Peng and A.~S. Parkins.
\newblock Motion-light parametric amplifier and entanglement distributor.
\newblock {\em Phys. Rev. A}, 65:062323, Jun 2002.

\bibitem{Gradshteyn2007}
I.~S. Grad{\v s}tejn and I.~M. Ry{\v z}ik, editors.
\newblock {\em Table of integrals, series, and products}.
\newblock Acad. Press, 7th. ed. edition, 2007.

\bibitem{Yurke2006}
B.~Yurke and E.~Buks.
\newblock Performance of cavity-parametric amplifiers, employing kerr
  nonlinearites, in the presence of two-photon loss.
\newblock {\em J. Lightwave Technol.}, 24:5054, Dec 2006.

\bibitem{Geroni2007}
M.~G. Genoni, M.~G.~A. Paris, and K.~Banaszek.
\newblock Measure of the non-gaussian character of a quantum state.
\newblock {\em Phys. Rev. A}, 76:042327, Oct 2007.

\bibitem{Kheruntsyan1999}
K.~V. Kheruntsyan.
\newblock {Wigner function for a driven anharmonic oscillator}.
\newblock {\em Journal of Optics B: Quantum and Semiclassical Optics}, 1:225,
  April 1999.

\bibitem{Gripp1995}
J.~Gripp, S.~L. Mielke, and L.~A. Orozco.
\newblock Cascaded optical cavities with two-level atoms: Steady state.
\newblock {\em Phys. Rev. A}, 51:4974, Jun 1995.

\bibitem{Stannigel2010}
K.~Stannigel, P.~Rabl, A.~S. S\o{}rensen, P.~Zoller, and M.~D. Lukin.
\newblock Optomechanical transducers for long-distance quantum communication.
\newblock {\em Phys. Rev. Lett.}, 105:220501, Nov 2010.

\bibitem{Anetsberger2009}
G.~Anetsberger, O.~Arcizet, Q.~P. Unterreithmeier, R.~Riviere, A.~Schliesser,
  E.~M. Weig, J.~P. Kotthaus, and T.~J. Kippenberg.
\newblock Near-field cavity optomechanics with nanomechanical oscillators.
\newblock {\em Nat Phys}, 5:909, 12 2009.

\bibitem{Astafiev2010}
O.~Astafiev, A.~M. Zagoskin, A.~A. Abdumalikov, Yu.~A. Pashkin, T.~Yamamoto,
  K.~Inomata, Y.~Nakamura, and J.~S. Tsai.
\newblock Resonance fluorescence of a single artificial atom.
\newblock {\em Science}, 327:840, 2010.

\bibitem{Akimov2007}
A.~V. Akimov, A.~Mukherjee, C.~L. Yu, D.~E. Chang, A.~S. Zibrov, P.~R. Hemmer,
  H.~Park, and M.~D. Lukin.
\newblock Generation of single optical plasmons in metallic nanowires coupled
  to quantum dots.
\newblock {\em Nature}, 450:402, Nov 2007.

\bibitem{SchoelkopfNature2008}
R.~J. Schoelkopf and S.~M. Girvin.
\newblock {\em Nature (London)}, 451:664, 2008.

\bibitem{Imamoglu1997}
A.~Imamo\ifmmode~\bar{g}\else \={g}\fi{}lu, H.~Schmidt, G.~Woods, and
  M.~Deutsch.
\newblock Strongly interacting photons in a nonlinear cavity.
\newblock {\em Phys. Rev. Lett.}, 79:1467, Aug 1997.

\bibitem{Hartmann2008}
M.~J. Hartmann, F.~G. S.~L. Brand{\~a}o, and M.~B. Plenio.
\newblock Quantum many-body phenomena in coupled cavity arrays.
\newblock {\em Laser \& Photonics Reviews}, 2:527, 2008.

\bibitem{Rebic2009}
S.~Rebi\ifmmode~\acute{c}\else \'{c}\fi{}, J.~Twamley, and G.~J. Milburn.
\newblock Giant {Kerr} nonlinearities in circuit quantum electrodynamics.
\newblock {\em Phys. Rev. Lett.}, 103:150503, Oct 2009.

\end{thebibliography}

\end{document}